\documentclass[conference]{IEEEtran}
\usepackage{bm}
\usepackage{amsmath}%
\usepackage{amssymb}
\usepackage{graphicx}

\usepackage{indentfirst}
\usepackage{url}
\usepackage{bm}

\ifCLASSOPTIONcompsoc
  \usepackage[nocompress]{cite}
\else
  \usepackage{cite}
\fi
\ifCLASSINFOpdf
\else
\fi
\usepackage[noend]{algpseudocode}

\usepackage{algorithmicx,algorithm}
\begin{document}
\title{Fair Auction and Trade Framework for Cloud VM Allocation based on Blockchain}

\author{\IEEEauthorblockN{Zhili Chen$^{1,2}$, Wei Ding$^{1,2}$, Yan Xu$^{1,2}$, Miaomiao Tian$^{1,2}$ and Hong Zhong$^{1,2}$}
\IEEEauthorblockA{Email: zlchen3@ustc.edu.cn, dwei109@126.com, xuyan@ahu.edu.cn, \\mtian@ahu.edu.cn, zhongh@mail.ustc.edu.cn}

1. School of Computer Science and Technology Anhui University Hefei, China \\
2. Anhui Engineering Laboratory of IoT Security Technologies, Anhui University, Hefei 230039, China

}



\maketitle

\begin{abstract}
Cloud auctions provide cost-effective strategies for cloud VM allocation. Most existing cloud auctions simply assume that the auctioneer is trustable, and thus the fairness of auctions can be easily achieved. However, in fact, such a trustable auctioneer may not exist, and the fairness is non-trivial to guarantee. In this work, for the first time, we propose a decentralized cloud VM auction and trade framework based on blockchain. We realize both auction fairness and trade fairness among participants (e.g., cloud provider and cloud users) in this system, which guarantees the interest of each party will not suffer any loss as long as it follows the protocol. Furthermore, we implement our system through the local blockchain and Ethereum official test blockchain, carry out experimental simulations, and demonstrate the feasibility of our system.
\end{abstract}

\begin{IEEEkeywords}
Blockchain, Cloud auction, Fair auction, Fair trade
\end{IEEEkeywords}

\IEEEpeerreviewmaketitle

\section{Introduction}

As the rapid development of cloud computing, combinatorial auctions are increasingly adopted for cloud VM allocation. Different from fixed price strategies which are widely used currently, cloud combinatorial auctions \cite{samimi2016combinatorial,baranwal2015fair,li2018online,zaman2013combinatorial} can reflect market supply and demand by providing dynamic price models, and meet the diverse requests (e.g., different types of cloud VMs) of users, such as CPU processing power, storage capacity, memory and so on.

Most existing cloud auctions \cite{samimi2016combinatorial,baranwal2015fair,li2018online,zaman2013combinatorial} implicitly assume that the auctioneer is trustable, and overlook the auction fairness. However, in practical applications, a trustable auctioneer may not exist. In such a case, some cloud users may collude with the auctioneer to learn bids of other users, and get more profit from the auction by taking advantage of this knowledge; or they may just quit the auction after learning other cloud users' bids. As a result, the \emph{auction fairness} could be violated.

Furthermore, after the auction, dishonest cloud users may choose not to trade cloud VM instances with the auctioneer for certain reasons (such as the lack of finance). They can simply abort the trade if there are no protection measures. More importantly, during the cloud trade, even if a cloud user has already paid, a dishonest service provider may not provide the requested service; or on the other hand, an honest provider may fail to get service fee when providing requested services to a dishonest users. In this way, the \emph{trade fairness} could also be violated.

The above analysis shows that, there is an urgent need to realize both auction fairness and trade fairness for cloud auction and trade systems, such that all participants should act honestly during cloud auction and trade, otherwise dishonest participants could be caught and punished. To this end, blockchain can be taken into account as an ideal means. As an emerging decentralized technology, blockchain provides transparent and credible data operations by cryptographic techniques. Due to its transparency and credibility, blockchain has been extensively studied and used to solve fairness issues in various fields, e.g., secure multiparty computations\cite{bentov2014use,andrychowicz2014secure,kumaresan2014use}, cloud computing\cite{zhang2018outsourcing,zhang2018blockchain,hu2018searching}, data trading\cite{delgado2017fair}, fog computing\cite{huang2018bitcoin} and so on. However, as far as we know, blockchain has not been applied to achieving both auction fairness and trade fairness in the context of cloud auctions.

In this paper, we propose a fair auction and trade framework for cloud VM allocation based on blockchain. We achieve both auction fairness and trade fairness by taking advantage of two key components of blockchain, i.e., cryptocurrency and smart contract. Specifically, we let cloud users commit to the blockchain with guaranties while bidding, and design a smart contract with verifiable states to compute both auction and trade processes. To ensure the auction fairness, the smart contract financially punishes dishonest participants, while compensates the honest ones immediately after the auction. Moreover, once the auction result comes out, the smart contract turns some parts of deposits into payments, and perform the cloud trade automatically. To further ensure the trade fairness, we design two trade schemes with and without an adjudicator, guaranteeing that honest cloud users pay only if they get required VM instances, while the honest provider provides required VM instances only if it gets the corresponding fees. Our contributions are as follows:
\begin{itemize}
  \item We combine cloud auction design with blockchain technology, and propose a fair auction and trade framework for cloud VM allocation. This framework is able to ensure both auction fairness and trade fairness with and without an adjudicator.

  \item To achieve auction fairness and trade fairness, we design a timed commitment scheme with guaranties, which financially punish dishonest participants, while compensates the honest ones according to their bids. Moreover, we design the state mechanism and ladder payment to guarantee trade fairness by verifying the validity of various states.

  \item We have fully implemented our design and carried out experiments on both simulated network and test network of Ethereum. Experimental results show the feasibility of our design.

\end{itemize}

The remainder of this paper is organized as follows. Section~\ref{sec:problem} is the problem definition. Preliminaries are given in Section~\ref{sec:preliminaries}. In Section~\ref{sec:challenges}, the state mechanism is described. In Section~\ref{sec:auction} and \ref{sec:withoutad}, we present our designs with and without an adjudicator in detail, and analyze their security, respectively. We briefly review the related work in Section~\ref{sec:relatedwork}. Then, in Section~\ref{sec:experiment} experiments are carried out. Finally, this paper is concluded in Section~\ref{sec:conclusion}.

\section{Problem Definition}\label{sec:problem}

\subsection{System model}

A combinatorial cloud auction setting is considered in our work, where a cloud provider supplies $m$ types of VM instances which is characterized by their sizes of the CPU processing power and the memory (e.g, small, medium, large and etc.), operating system (e.g., Linux, Microsoft Windows and etc.) and the time limit for VM instance usage (e.g., the number of hours that virtual machine instances can be used). For every instance resource type $i$, the provider can offer up to quantity $k_i$. For each user $u_j$ who wants to obtain cloud services needs to pay a fee and submit bid $B_{j} = \{(k^{1}_j,...,k^{m}_j),b_{j}\}$ consists of the number of VMs ${(k^{1}_j,...,k^{m}_j)}$ that $u_j$ wants to request and the price $b_{j}$ that $u_j$ is willing to pay, where $m$ denotes the number of VM types.

\subsection{Auction model}

The auction model proposed by Wang et al.\cite{wang2012cloud} solved the above cloud VM allocation and pricing problem without considering fairness. We sketch the two major phases of the auction model as follows.

{\bfseries(1) Resource allocation:} After collecting bids of users, the auctioneer computes
\begin{equation}\label{eq:density}
d_j = \frac{b_j}{\sqrt{\sum_{i=1}^m{k^i_j w_i}}} \qquad   1 \leq j \leq n
\end{equation}
for each user $u_j$, where $w_i$ denotes the weight of VM instances of type $i$. After all the bid densities are calculated, all users are ranked in descending order of bid densities.


Then the auctioneer traverses user list ordered previously, checks whether there are sufficient VM instances meeting the demand of each user $u_j$, i.e.,

\begin{equation}
\forall i \in \{1,...,m\}, \qquad \sum^{j-1}_{t=1}{k^i_tx_t}+k^i_j \leq k_i.
\end{equation}
and allocates the VM instances accordingly, where $x_t$ indicates whether user $t$ is allocated successfully.

{\bfseries(2) Payment scheme:} The auctioneer uses the following pricing rules:
\begin{itemize}
  \item A denied user pays 0;
  \item A user allocated with requested VM instances pays the \emph{critical value};
  \item A user having no \emph{critical user} pays 0.
\end{itemize}
where the $cirtical\;user$ of $u_j$ denotes the first denied user who would become winning without $u_j$ and the \emph{critical value} denotes the least bid value that user has to bid to win the auction. If $u_s$ is the \emph{critical user} of $u_j$, the auctioneer charges user $u_j$ as follows:
\begin{equation}
P_j = d_s \sqrt{\sum^{m}_{i=1}k^i_jw_i},
\end{equation}
where $P_j$ denotes the price that $u_j$ needs to pay.

\subsection{Adversary model and design goals}
Our concern is that if there is no trustable auctioneer, how can we achieve fairness in both the auction phase and the trade phase for cloud VM allocation. A lot of questions may arise without a trustable auctioneer. In the auction phase, a malicious user may collude with the auctioneer and modify its bid to pay a low price for the requested services, or directly quits the auction if not satisfied with the auction results known in advance. A malicious auctioneer may change the auction results for increasing its profit. In the trade phase, a malicious provider may reject to offer cloud services according to the auction outcome, even after getting a user's payment. On the other hand, a malicious user may get cloud services while refusing to pay the fee.

In our work, our design goals mainly consist of auction fairness and trade fairness as follows:

\begin{itemize}

  \item{Auction fairness:} The auction fairness means that it is infeasible for malicious users to alter their bids once submitted or to quit the auction without any fund punishment, and there is no chance for anyone to alter the auction outcome.

  \item{Trade fairness:} The trade fairness means that the trade is automatically executed once the auction is finished, and malicious users cannot get cloud VM instances without paying fees, while the malicious provider cannot get fees without providing requested VM instances.

\end{itemize}

 In summary, the main goal of fairness is to guarantee no matter how a malicious party behaves, the interest of a honest party will not suffer any loss.
\section{Preliminaries}\label{sec:preliminaries}

\subsection{Blockchain}
  Blockchain is a newly emerging decentralized technology which is transparent and credible. Blockchain is maintained by many miners jointly. These miners add new blocks in the blockchain by solving cryptographic problems. The hash of the previous block is embedded in the current block, the hash of the current block will also be embedded in the next block and so on. Therefore, the anti-tamper feature of blockchain is achieved by the mechanism that the previous block is unable to be altered unless alter this block and all the following blocks. Based on the anti-tamper characteristic, the credible nature of the blockchain is validated due to that the content of blockchain is approved by the collective voting of the miners. Besides, for the two current largest blockchain platforms: Bitcoin\cite{nakamoto2008bitcoin} and Ethereum\cite{Gavin2014ethereum,buterin2014next}, there is no threshold for becoming a miner. Hence, everyone is able to access the data without encryption of blockchain which guarantees the transparent feature of blockchain.

  Furthermore, blockchain has derived two excellent technologies: smart contract and cryptocurrency. In short, smart contract is the program on the blockchain that is uniquely specified by address on the blockchain. In virtue of blockchain, all the computations performed by a smart contract is believable and transparent. Cryptocurrency gets rid of central trustful agents and let participants transact directly without a bank.

  We use Ethereum as our platform, and for simplicity, we will not consider transaction fee. Note that \emph{msg.sender} in Ethereum is the address of user who sends this transaction or performs this operation.

  \subsection{Timed Commitment Scheme}

 Andrychowicz et al.\cite{andrychowicz2014secure} design a timed commitment scheme between a committer and a receiver to solve the problem of fair exchange of information in multi-party computation. In the scheme, the committer computes the commitment $h$ to a secret $s$ and posts $h$ along with the mutually predefined guaranty to the blockchain. The committer has to open the commitment by posting the secret $s$ on blockchain before a certain time to get its guaranty back. Otherwise the guaranty will flow to the receiver. Concretely, the timed commitment scheme consists of two phases:
  \begin{itemize}
    \item $commitment\;phase:$ The committer computes $h = H(s||r)$, where $s$ is the secret, $r$ is a random number, and $H(.)$ is a hash function. The commitment then posts the commitment $h$ along with a mutually predefined guaranty to the blockchain.
    \item $opening\;phase:$ The committer has to open commitment by posting the secret $s$ and randomness $r$ on the blockchain before a certain time to get its guaranty back. Otherwise the guaranty will flow to the receiver.
  \end{itemize}

 We adapt this timed commitment scheme to our cloud auction context, achieving the fairness with a financial penalty.

\section{State mechanism}\label{sec:challenges}

The \emph{timed commitment scheme} in the works\cite{bentov2014use,andrychowicz2014secure} is based on Bitcoin. Due to limited programmable capability of Bitcoin, the commitment scheme can only punish malicious behaviors breaking the deadline or providing wrong secrets. It is necessary for us to design a general mechanism that can be applied to more complex environments and easy implementation on Ethereum.

The Bitcoin script is simple and stateless while the Ethereum programming language (e.g., Solidity) is stateful and powerful which also brings new challenges. For example, it is impossible for users to repeat opening commitment in the Bitcoin script to get a guaranty twice, while it is feasible in the Ethereum platform to repeatedly obtain a guaranty due to that the smart contract can receive guaranties from multiple parties. The mechanism should make it impossible for users who have already disclosed their secrets to get a guaranty again by revealing their secrets.

We propose our \emph{state mechanism} to enure the conditions on which a certain operation is allowed. We use a tuple $(sid, Addr_{SC}, Addr_j, \delta_{SC}, \delta_{j}, \tau, \Omega_j, \Phi)$ to represent the environment state of an operation by a participant on the blockchain, where $sid$ denotes the session identifier for reusing the code of $SC$, $SC$ denotes the smart contract, $Addr_{SC}$ is the address of the smart contract, $Addr_j$ is the address of a user or the provider, $\delta_{SC}$ is the state of the smart contract, $\delta_{j}$ is the state of the user or the provider, $\tau$ is the deadline for performing this operation, $\Omega_j$ denotes the extra information (such as bid, guaranty, deposit and etc.) and $\Phi$ denotes the conditions that should be satisfied for completing this operation.

 As time passes e.g., the deadline $\tau$ switches, the smart contract enters a different stage, and $\delta_{SC}$ changes accordingly. Similarity, $\delta_{j}$ changes as the participant $j$ makes some behavior. Note that it is easy to implement various states for participants and the smart contract $SC$ with enumeration type in the Solidity. Moreover, $\Phi$ plays an important role in judging malicious acts. A behavior is valid if and only if the following conditions
\begin{equation}\label{eq:state}
  \begin{split}
  & Addr_j=msg.sender\\
  & \Phi_{SC} = \delta_{SC}\\
  & \Phi_{j} = \delta_{j}\\
  & \Phi_\Omega \asymp \Omega_j\\
  \end{split}
\end{equation}
are all satisfied. Here, the former three equations are basic conditions ensuring the right participant ($msg.sender$) has the right state ($\Phi_{j}$) to send a transaction at the right time (same as $\Phi_{SC}$), while the last equation denotes the other conditions ($\Phi_\Omega$) that should be satisfied (such as the sent bid is hashed and the answer is equal to the commitment value), where $\asymp$ denotes extra information should satisfy the extra conditions. Furthermore, with the help of dictionary type, we map $Addr_j$ to other information belong to $u_j$ such as $\delta_j$, $\Omega_j$ and etc. Hence, the other information belongs to $u_j$ is bound to $Addr_j$ to make sure that only $u_j$ himself is able to operate on his own data. The former two conditions merely depends on the correctness of the smart contract. As long as the smart contract $SC$ is well programmed, these two equations are naturally satisfied. Unlike the previous two equations, whether the latter two equations are satisfied is determined by the behavior and current state of participant $j$. Therefore, it is necessary to consider the last two conditions e.g., the behaviors and the states of users or the provider to determine whether a blockchain operation is allowed. Finally, we simplify the \emph{state mechanism} to the two equations as follows.

\begin{equation}
  \begin{split}
  &\Phi_{j} = \delta_{j}\\
  &\Phi_\Omega \asymp \Omega_j  \\
  \end{split}
\end{equation}

In summary, our \emph{state mechanism} guarantees that the right participant should do the right thing at the right time. Otherwise, the participant will be punished by a predefined monetary penalty or the operation will be rejected.

\section{Our Framework}\label{sec:auction}

In this section, we give the framework overview, present our framework with an adjudicator in detail, and analyze its security.

\subsection{Framework Overview}

\begin{figure}
\centering
\includegraphics[scale=0.4]{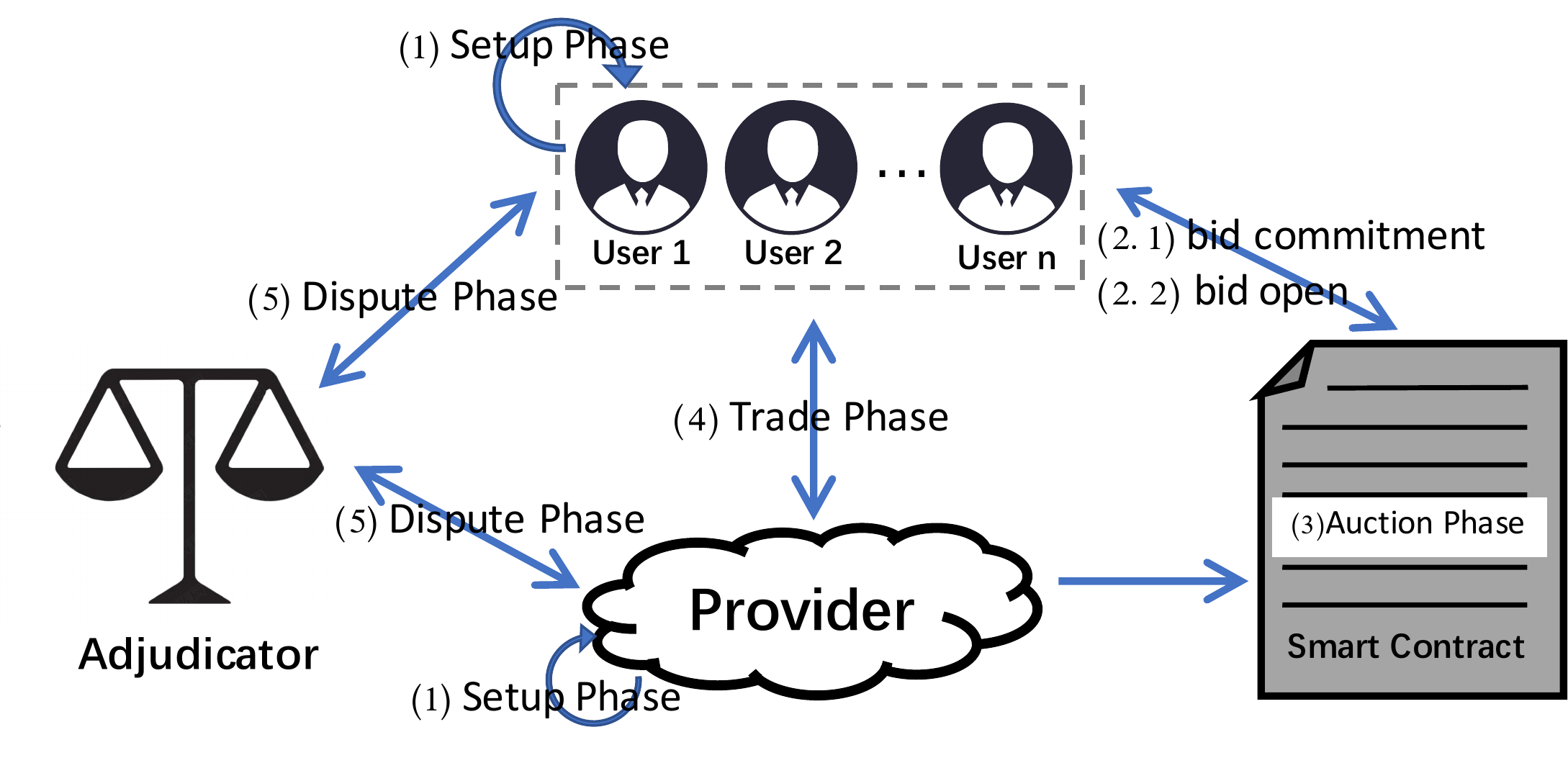}
\caption{The interactions between parties of the framework}
\label{fig:label1}
\end{figure}

Our main idea is to simulate a trustable auctioneer with a smart contract (SC), and make use of cryptocurrency for achieving both auction fairness and trade fairness. Moreover, we use an adjudicator to authenticate the validity of cloud VM instances. The flow chart of our framework is illustrated in Fig.~\ref{fig:label1}. There are five phases as follows.

  \textbf{(1)Setup phase.} The participants generate their own parameters like public-private key pairs, and prepare digital coins for their bid commitment guarantees.


  \textbf{(2)Bid phase.} The participants commit their bids through the timed commitment, and then all commitments are opened to the smart contract, and the corresponding deposits are made.

  \textbf{(3)Auction phase.} With the bids submitted, the smart contract computes the auction, allocates the cloud VM instances to cloud users, and prices accordingly.

  \textbf{(4)Trade phase.} The smart contract executes the trade automatically, such that the assess permissions of cloud VM instances are given to users, and the deposits of corresponding users are transferred to the provider if there is no dispute.

  \textbf{(5)Dispute phase.} When some disputes happened in the trade, participants are able to solve the disputes with the help of the adjudicator.


  \subsection{Detailed Construction}

  Now, we describe our framework in detail. For simplicity, we denote a transaction by $TX$ and its amount of money by $|TX|$. For two long-term parties, the provider and the adjudicator, their addresses (e.g., $Addr_p$ and $Addr_A$) are stored in the smart contract before all auctions begin.

\begin{figure}[t]
  \centering
    \includegraphics[scale=0.4]{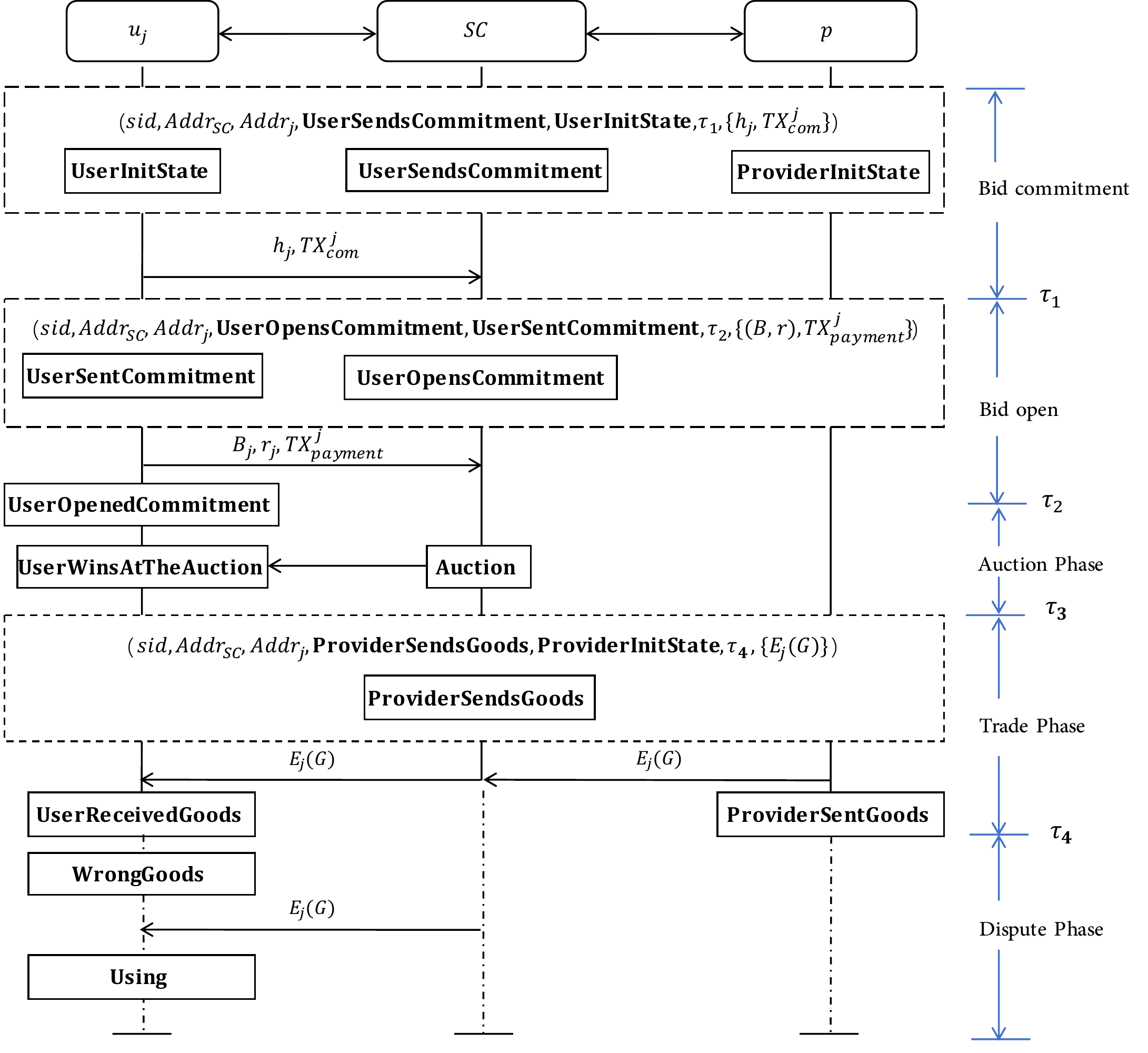}
  \caption{The overall state changes of the system}
  \label{fig:label2}
\end{figure}

    \subsubsection{Setup phase}

    In this phase, participants generate some system parameters. Specifically, each user $u_j$ generates a ECDSA key pair $(pk_{j},sk_{j})$, and then its address $Addr_j$ is computed by hashing its public key $pk_j$. It is preferred that a user recreates a new pair of keys each time it participates in the auction, for maintaining anonymity and protecting privacy.  The provider transfers a deposit to the smart contract to ensure its correct behaviors. The deposit value is proportional to the total weighted number of VM instances the provider provides, and the deposit value per weight unit is defined as base price $\beta$. Each user $u_j$ transfers a balance of value $a$ which is locked by its own secret key $sk_j$. And the balance is used as a guaranty for the timed commitment. For a new $sid$, $\delta_j$ and $\delta_p$ are currently in the states of \textbf{UserInitState} and \textbf{ProviderInitState}, respectively.

\begin{algorithm}[t]
\caption{BID PHASE}
\label{algorithm1}
  \begin{algorithmic}[1]
    \State \underline{{\bfseries procedure} BidCommitment}
    \For{$j=1,...,n$}
      \State // Function \textbf{require} aborts execution and reverts state changes if the condition is not met
      \State \textbf{require}($\delta_{SC}$==\textbf{UserSendsCommitment})
      \State \textbf{require}($\delta_j$==\textbf{UserInitState})
      \State Receive $h_j$ and $TX^j_{com}$ from user $u_j$ before $\tau_1$
      \If{$h_j$ $\neq$ NULL $\&\&$  $|TX^{j}_{com}|==a$}
      \State $\delta_j$ = \textbf{UserSentCommitment}
      \EndIf
    \EndFor
    \If{(All users have committed) OR (Time $\tau_1$ is past)}
    \State $\delta_{SC}$ = \textbf{UserOpensCommitment}
    \EndIf
    \\
    \State \underline{{\bfseries procedure} {BidOpen}}
    \For{$j=1,...,n$}
      \State \textbf{require}($\delta_{SC}$==\textbf{UserOpensCommitment})
      \State \textbf{require}($\delta_j$==\textbf{UserSentCommitment})
      \State Receive $B_j$, $r_j$, $TX^{j}_{payment}$ from user $u_j$ before $\tau_2$

      \If{$h_{j}$==$H(B_{j},r_j, Addr_j)$ $\&\&$ $|TX^{j}_{payment}|$==$b_{j}$}
        \State $\delta_j=\textbf{UserOpenedCommitment}$
        \State $d_j = \frac{b_j}{\sqrt{\sum_{i=1}^m{k^i_j w_i}}}$
      \Else
        \State $\delta_j = \textbf{UserFailsToOpenCommitment}$
      \EndIf
    \EndFor
    \For{$j=1,...,n$}
      \If{$\delta_j$= \textbf{UserOpenedCommitment}}
      \State $SC \rightarrow u_j: TX^{j}_{refund}$
      \EndIf
    \EndFor
    \If{(All commitments is opened) OR (Time $\tau_2$ is past)}
    \State $\delta_{SC}$ = \textbf{Auction}
    \EndIf
  \end{algorithmic}
\end{algorithm}

    \subsubsection{Bid phase}

To submit their bids simultaneously, the users first commit their bids and then open the comments to the smart contract. Thus, this phase comprises the following two steps.

 {\bfseries Bid commitment.} At this step, the smart contract $SC$ requires users to send their commitments and guaranties to it before $\tau_1$. A user will not be able to join the current auction if it fails to do the submission. At this time, $\delta_{SC}$ and $\delta_j$ of user $u_j$ are \textbf{UserSendsCommitment} and \textbf{UserInitState}, respectively.

Without loss of generality, we assume that there are a total of $n$ users involved. Specifically, each user $u_j$, for $1 \leq j \leq n$, computes the hash value $h_j = H(B_{j}||r_{j}||Addr_j||sid)$ locally, and posts $h_{j}$ with a guaranty value $a$ on $SC$ before a certain time $\tau_1$. In consideration of the small range of $B_{j}$ values, a uniform random number $r_j \in \{0,1\}^\lambda$ is also involved while computing the hash value to ensure security, where $\lambda$ is the security parameter. Besides, the parameters $Addr_j$ and $sid$ indicate that the commitment is from user $u_j$ and the ID of the current auction is $sid$, meaning that only user $u_j$ can disclose this commitment at the auction $sid$. For Ethereum, user $u_j$ calculates $h_j$ as follows:
\begin{equation}
h_j=keccak256(B_j,r_j,Addr_j, sid)
\end{equation}
 and posts $h_{j}$ with guaranty to $SC$ before $\tau_1$, and then $\delta_{j}$ will become \textbf{UserSentCommitment}.

The commitment step is shown in Algorithm \ref{algorithm1}. A user with the \textbf{UserInitState} state satisfies the basic condition. The condition of \textbf{UserInitState} state guarantees that the address (thus the corresponding user) that has not attended before is able to participate in the current auction with $sid$, and all addresses that have been committed cannot be used again. Besides, if the extra conditions of non-empty commitment and the guaranty value being $a$ are satisfied, $\delta_{j}$ will become \textbf{UserSentCommitment}. At the end of bid commitment step, the state of $SC$ becomes \textbf{UserOpensCommitment} if all users have committed or time $\tau_1$ is past.

For simplicity, we assume that all $n$ users have joined the auction, that is, every user sends the commitment with the guaranty to $SC$ before $\tau_1$. The procedure of bid commitment is also shown in Figure $\ref{fig:label2}$.

{\bfseries Bid open.} At the step, $SC$ with \textbf{UserOpensCommitment} requires users with \textbf{UserSentCommitment} state to open their commitments with deposits before $\tau_2$ ($\tau_2 > \tau_1$), where deposits amount to their bid values and are sufficient to pay charges to the provider. A user who fails to open its commitment before $\tau_2$ will pay the penalty with its guaranty. Note that we regard this kind of users as quitting the auction. This step is demonstrated in Figure $\ref{fig:label2}$.

Each user $u_j$ sends $B_{j}$ and $r_{j}$, and transfers a deposit of value $b_{j}$ to the smart contract $SC$, then $\delta_j$ will turn into \textbf{UserOpenedCommitment} state. Note that transferring deposits to $SC$ is to ensure that the provider $p$ gets the corresponding fees as long as it provides the right VM instances. The users are able to withdraw their deposits if the auction phase fails or the provider provides unexpected VM instances. Hence, their interests will be well protected.

The basic condition of each user $u_j$ will be met, if $\delta_j$ is \textbf{UserSentCommitment}. Extra conditions require users $u_j$ to open their commitments correctly (e.g., sending correct $B_j$ and $r_j$) with deposit values $b_j$. If all the conditions are satisfied, $\delta_j$ will become \textbf{UserOpenedCommitment}. Otherwise, it is impossible for users $u_j$ to get their guaranties back and $\delta_j$ will be \textbf{UserFailsToOpenCommitment}. Furthermore, once the users have successfully opened commitments and withdrawn their guaranties, it is impossible for them to withdraw the guaranties again due to the change of state.

Once comments are opened, refunds are handled for users. There are three cases: (1) No refund is made for users with \textbf{UserFailsToOpenCommitment} state; (2) Only guaranties $a$ are returned to users with \textbf{UserOpenedCommitment} state but with bid densities $d_i < \beta$; (3) The refunds consisting the guaranties and the compensation are made for users with \textbf{UserOpenedCommitment} state and with bid densities $d_i \ge \beta$. The refunds of the third case can be computed by Eq.~\eqref{equation8}, where $n_f$ denotes the number of users of the first case, $\xi_j = (d_j>=\beta)$ indicates whether the bid density is no less than the base price $\beta$.
  \begin{equation}
  \begin{split}
|TX^{j}_{refund}| =  n_f \cdot a \cdot \frac{\xi_j \cdot d_{j}}{\sum_{y=1}^{n}{(\xi_y \cdot d_{y})}} + a
  \label{equation8}
  \end{split}
\end{equation}
Here, we use a non-uniform compensation in the auction, where compensation is only made for users with bid densities no less than $\beta$, and the compensation values grow proportionally with bid densities.

Note that since the Ethereum requires that the execution of a smart contract must be triggered by an external user-controlled account, the smart contract $SC$ cannot automatically run the auction, even if $\tau_2$ has arrived, unless all $n$ users disclose their secrets and trigger the critical condition. Hence, a participant who always activates the smart contract to execute the auction phase is necessary. In the light of the long-term participation and the drive of interest, the provider is the first choice. Meanwhile, to prevent the malicious provider from not activating the smart contract and causing user payments to be trapped, another deadline $\tau_3(\tau_3>\tau_2)$ is introduced to protect the interests of users. If the auction phase does not begin, but time $\tau_3$ is reached, then the users are able to get their own payments back.

  \subsubsection{Auction phase}

  In this phase, the smart contract $SC$ runs a cloud auction, determines winning users and their payments. Different from the previous work \cite{wang2012cloud}, we use a smart contract to replace the trustable auctioneer who is often played by the provider. Due to the transparency and credibility of smart contracts, the auction execution based on a smart contract is trustable, as long as the smart contract is correctly prepared. The smart contract adapts auction mechanism essentially the same as that of \cite{wang2012cloud} except with appropriate state changes, as shown in Algorithm \ref{algorithm2}. In the algorithm, Lines 1-9 determine winners, set the state of the winners as \textbf{UserWinsAtTheAuction}, and that of the losers as \textbf{UserFailsInTheAuction}. Lines 10-18, compute prices for the winners, and give back payments to all losers.

    \begin{algorithm}[t]
  \caption{AUCTION PHASE}
  \label{algorithm2}
  \begin{algorithmic}[1]
  \State {// Resource allocation}
  \State $SC$ sort $d_j$ such that $d_1 \geq ... \geq d_N$
  \For{j=1,...,N}
    \If{$\forall$ $i$ $\in$ $\{1,...,m\}$, $\quad$ $\sum^{j-1}_{t=1}{k^i_tx_t}+k^i_j \leq k_i$}
    \State $x_j=1$
    \State $\delta_j=\textbf{UserWinsAtTheAuction}$
    \Else
    \State $x_j=0$
    \State $\delta_j=\textbf{UserFailsInTheAuction}$
    \EndIf
  \EndFor
  \State {// Payment scheme}
  \For{j=1,...,N}
  \If{$\delta_j=\textbf{UserWinsAtTheAuction}$}
  \State $\overline{ins_1}=\sum^{j-1}_{t=1}k^1_tx_t ,..., \overline{ins_m}=\sum^{j-1}_{t=1}k^m_tx_t$
  \For{s=j+1,...,N}
  \If{$\forall q \in\{1,...,m\},\overline{ins_q}+k^q_s \leq k_q$}
  \State $\overline{ins_1}=\overline{ins_1},...,\overline{ins_m}=\overline{ins_m}+k^m_s$
  \If{$\exists q \in \{1,...,m\},\overline{ins_q}+k^q_j > k_q$}
  \State $P_j=d_s\sqrt{\sum^{m}_{i=1}{k^i_jw_i}}$
  \EndIf
  \EndIf
  \EndFor
  \EndIf
  \EndFor
  \State $\delta_{SC}$ = \textbf{ProviderSendsGoods}
  \State Give back payments to all losers
  \end{algorithmic}
  \end{algorithm}

\subsubsection{Trade phase}

In the phase, the smart contract $SC$ requires the provider to send encrypted permission information of VM instances $E_j(G) = E_{pk_j}(G)$ to winners $u_j$ before $\tau_4$ ($\tau_4 > \tau_3$). If the provider rejects to send the permission information, as soon as time $\tau_4$ is reached, all winners will get their payments back, and also get deposit values (i.e., $\beta \cdot \sum_{i=1}^m{k^i_jw_i}$) transferred by the provider as compensations. Otherwise, $\delta_p$ becomes \textbf{ProviderSentGoods} state after the provider has sent $E_j(G)$, and $\delta_j$ becomes \textbf{UserReveivedGoods} state if winner $j$ gets $E_j(G)$ from $SC$. After this, winers $j$ check whether the VM instances satisfy their requirements. If not, the trade phase will jump to the dispute phase. If there is no problem about the VM instances after all winners have used the VM instances, the provider will get the payments.





\subsubsection{Dispute phase}

In the phase, the adjudicator deals with disputes for the provider and the winners. If winners $u_j$ get unexpected VM instances, they will appeal to the adjudicator for help, and $\delta_j$ will become \textbf{WrongGoods}. The adjudicator checks whether $\delta_p$ is \textbf{ProviderSentGoods}, and requires the provider to send the adjudicator the encrypted permission information of qualified VM instances $E_{pk_A}(G)$ through $SC$, where $pk_A$ is the public key of the adjudicator. Then the adjudicator checks whether the VM instances meet the requirements of winners $u_j$. If not, the provider will be considered malicious and the adjudicator will help winners $u_j$ get payments back, and get the compensation as described in the trade phase. Note that the address of the adjudicator is written when the smart contract is deployed. With this unique address, the adjudicator can perform a withdrawal function to help the winners get their payments back. If the VM instances are qualified, the adjudicator will send $E_j(G)$ to winners $u_j$ and $\delta_j$ will be \textbf{Using}. Even if the winners $u_j$ are malicious, there is no harm to the provider for the provider just needs to send the valid VM instances to the adjudicator again.

\subsection{Security Analysis}

{\noindent \bfseries Theorem.} Our framework satisfies auction fairness if the hash function $H$ is collision-resistant.

{ \noindent \bfseries Proof.} In our context, auction fairness means that it is infeasible for malicious users to alter their bids once committed, or to quit the auction without financial penalties, and it is infeasible to change the auction result once all bids are opened. We examine these issues as follows.


(1) We show that it is infeasible for malicious users to alter their bids once committed. This is quite straightforward due to the collision-resistance of the hash function $H$. Specifically, if a malicious user $u_j$ wants to alter its bid $B_j$ to a different value $B'_j$ after the commitment, it must find an appropriate $r'_j$, such that $h_j = H(B_j||r_j||Addr_j||sid)=H(B'_{j}||r'_j||Addr_j||sid)$. This means that the user $u_j$ manages to find a collusion for function $H$, which conflicts with the collision-resistance assumption.


(2) We show that malicious users cannot quit the auction without any financial penalties. Once a malicious user $u_j$ commits to its bid in the bid phase, it participates in the auction. During the commitment, the user $u_j$ must also post a guaranty to the smart contract. The only chance that the user $u_j$ can quit the auction is when opening the bid. Then, if the user $u_j$ does not open its bid, it will be punished with its guaranty $a$; otherwise, if opening the commitment, the user $u_j$ must also send a deposit of its bid value to the smart contract, and it has no other chance to quit until the the auction is finished.


(3) We show that the auction result cannot be changed once all bids are opened. The smart contract emulates a trustable auctioneer. Once all the bids are opened, the auction result will be determined, since the smart contract will execute the auction honestly. Our state mechanism defines valid operations for participants, guarantees the correctness of the auction result, and thus there is no chance for any participant to change the auction result.

Conclusively, our framework satisfies the auction fairness.
\\$\Box$

{\noindent \bfseries Theorem.} Our framework satisfies the trade fairness.

{ \noindent \bfseries Proof.} In our context, the trade fairness means that it is impossible for the malicious provider to reject trade without any financial penalty after learning the auction result or to get payments without offering VM instances, and it is impossible for malicious users to stop trading without any financial penalty after learning the auction result, or to get valid VM instances without paying fees. We will prove the trade fairness in these two aspects as follow.

\emph{Case 1: The provider is malicious and a user $u_j$ is honest.} In this case, if the provider rejects to trade, then it will lose its deposit transferred in the setup phase, and the user $u_j$ gets its compensation. If the provider provides unexpected VM instances to get the payment, the user $u_i$ would find the problem and resort to the adjudicator, who would supervise the provider to send valid VM instances before time $\tau_5$. If the provider still fails to provide valid VM instances before time $\tau_5$, the adjudicator would help $u_j$ get its payment back, and get part of the provider's deposit as compensation. Therefore, the trade fairness holds in this case.

\emph{Case 2: A user $u_j$ is malicious and the provider is honest.} In this case, if the user $u_j$ stops trading, it would lose its payment, and the provider get the payment as its compensation. If the user $u_j$ wants to get valid VM instances without paying to the provider, it must resort to the adjudicator to get its deposit back. Nevertheless, the adjudicator is trustable and would judge the VM instances are valid. As a result, the deposit of the user $u_j$ would be automatically transferred to the provider as the payment. Thus, the trade fairness holds also in this case.

In summary, our framework satisfies the trade fairness.
\\$\Box$

\section{Trade without an Adjudicator}\label{sec:withoutad}

Sometimes it is hard to find an adjudicator. For example, the users and the provider come from different countries, and it is difficult to find an adjudicator to resolve cross-border disputes. In this case, without an adjudicator, things become much worse and complicated. The provider has absolute control over its resources, it may get payments without providing valid VM instances due to there is no adjudicator who helps users to resolve a dispute. Hence, we relax the requirement of the trade fairness.

\subsection{Trade Fairness without an Adjudicator}

Due to the provider's absolute control over its resources, users are in inferior positions in the cloud trade. Hence, we propose the trade fairness without an adjudicator that allows the provider to tolerate a tiny loss less than or equal to $P_{tolerate}$. This trade fairness ensures that the provider gets corresponding service fees as long as it offers valid VM instances, and the users gets their VM instances for the required periods of time as long as they pay the corresponding service fees, both except for a controllable tiny loss.

\subsection{Ladder Payment}

In the trade with an adjudicator, users will lose all deposits without the help of the adjudicator if the provider simply shuts down VM instances. This will do a great harm to users. We find that it is possible to reduce economic losses by splitting the service time, such that participants will only suffer partial losses.

In practical applications, we split usage time into time segments and let users send a confirmation transaction to the smart contract $SC$ to confirm the use of each time segment.  The trade process proceeds as follows. The provider first sends VM instances to users, allowing them to use a time segment, then  the users must send the confirmation transaction to $SC$ before they resume the second time segment. This process repeats until the use time reached. Finally, the smart contract $SC$ pays the provider with the amount proportional to the total number of confirmed time segments.

\begin{figure}
\centering
\includegraphics[scale=0.45]{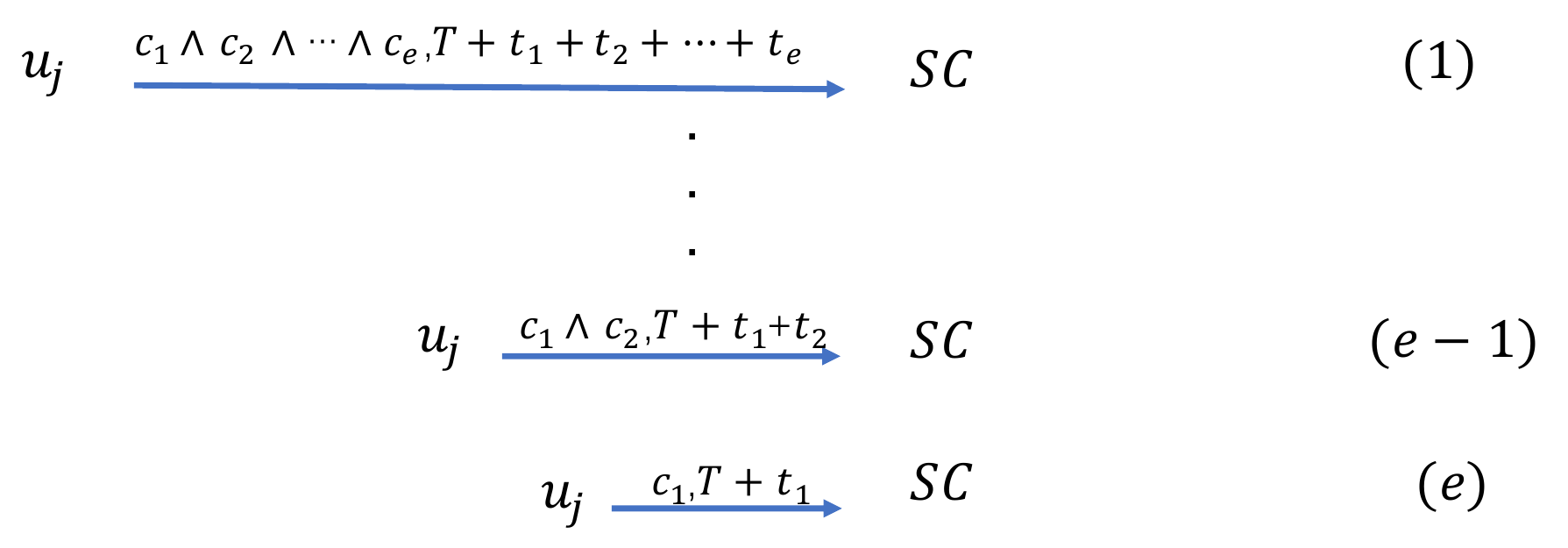}
\caption{The \emph{ladder payment mechanism}}
\label{fig:label5}
\end{figure}

Furthermore, due to the continuity of usage times, it is necessary to ensure that for the confirmations of two adjacent time segments, one follows another. Inspired by the ladder mechanism in \cite{bentov2014use}, we propose our \emph{ladder payment mechanism} as illustrated in Figure $\ref{fig:label5}$,  where $e$ denotes the total number of time segments, $T$ denotes the time that the user receives the cloud VM instances, $c_i$ denotes $i$th confirmation, $t_i$ denotes $i$th usage time and $\xrightarrow{c_1 \wedge c_2 \wedge... \wedge c_i,T+t_1+t_2+...+t_i}$ denotes it is essential for users to complete all the $i-1$ previous confirmations and send the current confirmation before $T+t_1+...+t_i$ if he wants to confirm the $i$th use time.

We can use the \emph{state mechanism} to simplify the latter payment mechanism. Let the initial state for using cloud VM instances be $\mathbf{Using_1}$, which means the user is currently using the $1$th segment of time. And $\mathbf{Using_1}$ will become $\mathbf{Using_2}$ after the confirmation $c_1$. Hence, once the user completed all $i-1$ previous confirmations and the state of user will become $\mathbf{Using_i}$. In this way, we can use the equation $(\ref{equation9})$ to demonstrate our \emph{ladder payment} where $T_i$ is equal to the sum of $T+t_1+...+t_i$.
\begin{equation}
\label{equation9}
u_j(Addr_j,\mathbf{Using_i})\xrightarrow{c_i,T_i}{SC(Addr_{SC},\mathbf{Using_i})}
\end{equation}

\begin{figure}
\centering
\includegraphics[scale=0.42]{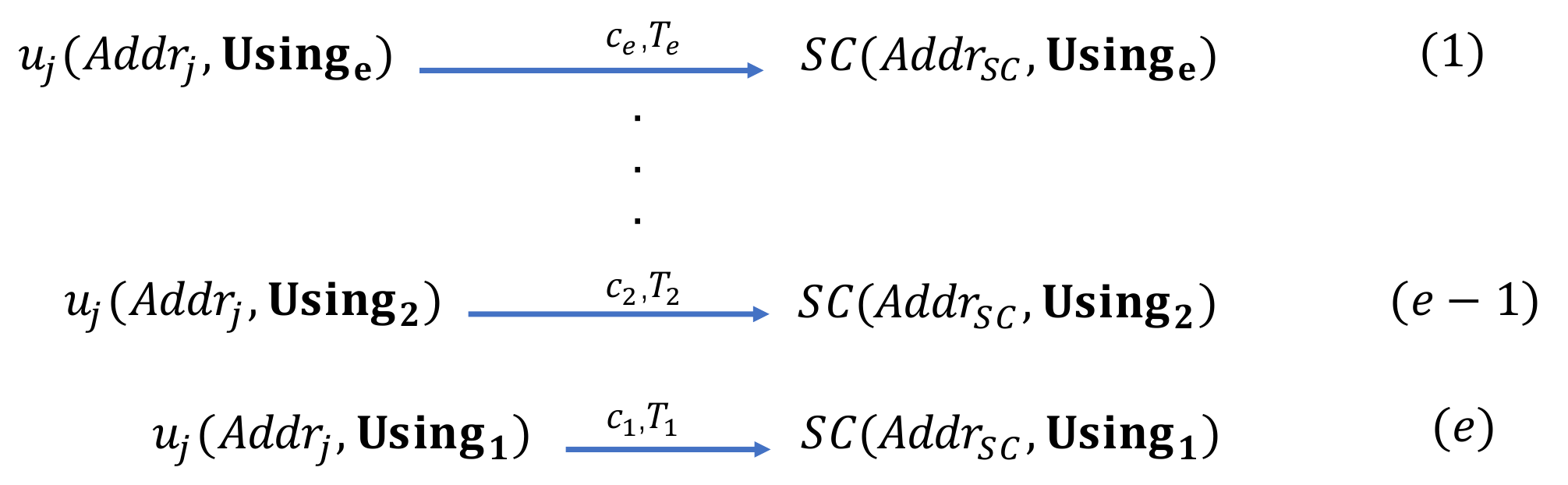}
\caption{The simplified \emph{ladder payment mechanism}}
\label{fig:label6}
\end{figure}

Figure $\ref{fig:label6}$ illustrates the simplified \emph{ladder payment}. It is not regarded malicious if $u_j$ does not send a correct confirmation in a specific state. For example, the user $u_j$ in $\mathbf{Using_1}$ must send the confirmation $c_1$, and other confirmations will be rejected. Actually, if the user $u_j$ refuses to send confirmation transactions, the provider $p$ will refuse to provide further time segments, and the trade ends.

\subsection{Our Framework without an Adjudicator}

We can modify our framework without an adjudicator with four phases: setup phase, bid phase, auction phase and trade phase. The first three phases are same as our framework aforementioned except that there is no adjudicator. The trade phase is quite different from that of the previous framework, and we detail it as follows.

With the help of the simplified \emph{ladder payment}, we design the trade phase without an adjudicator.
The usage time of VM instances is divided into $e$ equal segments. Hence, each time segment is of length $\frac{\widetilde{T}}{e}$, and $T_i=\tau_4+\frac{i\cdot \widetilde{T}}{e}$. The provider $p$ trades with each user $u_j$ as follows.

   { \bfseries(1)} After auction phase, the state of the user $u_j$ $\delta_j$ is $\textbf{UserWinsAtTheAuction}$, and the provider $p$ sends $E_j(G)$ before $\tau_4$ and $\delta_j$ will become $\mathbf{Using_1}$.

  {\bfseries(2)} Upon receiving the VM instances from $p$, $u_j$ checks whether they meet the requirements. If not, $u_j$ posts a disaffirmation $\overline{c}$, or just refuses to send any confirmation to $SC$. And $p$ will prevents $u_j$ from using the VM instances. Then the deposit of $u_j$ will flow to $u_j$ after $u_j$ activates ${SC}$. Note that the disaffirmation $\overline{c}$ differs from confirmation that $\overline{c}$ will only be sent once, thus the $\overline{c}$ is same. If the requirements of $u_j$ are satisfied, $u_j$ will use cloud VM instances for its purposes.

  {\bfseries(3)} After using the time segment $i$,  the user $u_j$ needs to post the confirmation $c_i$ to $SC$ for further use, and $\delta_j$ of $\mathbf{Using_{i}}$ state will turn into the state of $\mathbf{Using_{i+1}}$. Otherwise, the provider $p$ will keep the user $u_j$ from its VM instances.

  {\bfseries(4)} After being aware of the new confirmation, the provider $p$ performs operations for its best interest. If the usage time of $u_j$ has run out, the provider $p$ will stop offering the user $u_j$ with corresponding VM instances, and put these VM instances into an auction again for more revenue. If there is still usage time left for the user $u_j$, the provider $p$ will maintain the present status, and let the user $u_j$ continue to use for more profit. Moreover, if the provider $p$ is malicious, and forbids the user $u_j$ from using VM instances, the user $u_j$ will not post confirmations to $SC$, and at the same time the provider $p$ will not gain more service fee from $u_j$.

  {\bfseries(5)} Steps 3) and 4) iteratively operate many times, until one party fails to continue or there is no usage time left.

  {\bfseries(6)} When the trade ends up, the smart contract $SC$ allocates the amount of deposit according to the number of $u_j$'s confirmations. The provider $p$ gets payment
  \begin{equation}
   |TX^{p}_{payment}|=\frac{i}{e} \cdot P_j.
   \end{equation}
   Meanwhile, the user $u_j$ gets back the remain of the deposit
   \begin{equation}
   |TX^{j}_{remain}|=P_j \cdot(1- \frac{i}{e}).
   \end{equation}

In summary, by splitting the usage time and posting confirmations to the smart contract $SC$, we can decrease the loss of a honest participant in the trade. Splitting the usage time reduces the risk of the trade from $P_j$ to $\frac{P_j}{e}$. Posting confirmations to $SC$ is a guarantee for both users and the provider. Specifically, a confirmation ensures that the provider receives the corresponding payment, while it also guarantees the interests of users by ensuring the provider provides valid VM instances.




\subsection{Security Proof}

{\noindent \bfseries Theorem.} The trade without an adjudicator satisfies the trade fairness as long as the total number of time segments $e$ is greater than or equal to $\frac{P_j}{P_{tolerate}}$.

{ \noindent \bfseries Proof.} The trade fairness without an adjudicator guarantees the provider gets service fees as long as it provides valid VM instances, and each user $u_j$ get valid VM instances as long as it pays the corresponding fee, except that a honest participant may occur a controllable loss. We prove the trade fairness as follow assuming there are $e$ time segments.

\emph{Case 1: If the provider $p$ is malicious and the user $u_j$ is honest.} In this case, $p$ aims to get service fees without sending valid VM instances to the user $u_j$. If $p$ sends invalid VM instances, $u_j$ will not send a confirmation to $SC$ after a short inspection, and $p$ gets no fee. If $p$ aims to get more payment without providing $u_j$ with a sufficient length of usage time, $u_j$ will only to send confirmations to $SC$ according to the valid time segments. Hence, the interest of $u_j$ will not be damaged and $p$ will not get more benefit.

\emph{Case 2: If a user $u_j$ is malicious and the provider $p$ is honest.} In this case, the user $u_j$ attempts to get the usage time of VM instances without paying to $p$. If $u_j$ denies to send a confirmation or sends a disaffirmation, $p$ will refuse to provide more usage time. Nevertheless, when $u_j$ denies to send a confirmation, $p$ has already provided the corresponding time segment, and the loss is unavoidable. Therefore, the interest of $p$ may suffer a loss of value $\frac{P_j}{e}$. The trade fairness is still guaranteed as long as $\frac{P_j}{e}$ is less than or equal to $P_{tolerate}$. Therefore, $e$ should be greater than or equal to $\frac{P_j}{P_{tolerate}}$.

In summary, the trade without an adjudicator also satisfies the fairness.

\section{Related Work}\label{sec:relatedwork}

In this section, we review the related work and distinguish our work from existing ones.

There exist many cloud auction schemes \cite{Amazon2018a,li2018online,zaman2013combinatorial,samimi2016combinatorial,wang2013reverse}, solving various cloud auction settings. Papers \cite{li2018online,zaman2013combinatorial} proposed combinatorial auction schemes, addressing the problem that \emph{Spot Instance}\cite{Amazon2018a} does not distinguish different cloud VMs. Meanwhile, Samimi et al.\cite{samimi2016combinatorial} designed a combinatorial double auction scheme supporting auctions with multiple users and multiple providers. To allocate various cloud VMs from different providers, Wang et al.\cite{wang2013reverse} proposed a reverse batch matching auction scheme. However, this line of work does not consider the fairness, which may hinder the widespread use of cloud auctions.

Some papers \cite{andrychowicz2014fair,andrychowicz2014secure,bentov2014use,kumaresan2014use} have used blockchain to solve fairness problems. Andrychowicz et al. proposed fair two-party secure computations \cite{andrychowicz2014fair}, and fair multi-party secure computations \cite{andrychowicz2014secure} with $O(n^2)$. Bentov et al.\cite{bentov2014use} and Kumaresan et al.\cite{kumaresan2014use,bentov2017instantaneous} demonstrated fair SMC with $O(n)$. Furthermore, papers \cite{delgado2017fair,huang2018bitcoin,zhang2018blockchain,zhang2018outsourcing,hu2018searching} solved various fairness problems based on blockchain. Nevertheless, there is little work based on the blockchain to solve auction fairness and trade fairness in the setting of cloud computing.

The most relevant works to ours are researches on fair payment in cloud computing. Papers \cite{huang2018bitcoin,zhang2018blockchain,zhang2018outsourcing} implement the trade fairness for cloud computing services based on verifiable computing. However, our aim is to solve the problem of VM trade in the physical world, which demonstrates more difficulties such as that VMs may be suddenly shut down or VMs may not meet the configuration requirements and and so on. The trade fairness in our context is more complex. Besides, we also take into account the auction fairness.

\section{Implementation and evaluations}\label{sec:experiment}

Our design was deployed on the local simulated network \emph{Ganache} and on the official Ethereum test network \emph{Rinkeby}, respectively. The \emph{Rinkeby} network simulates the working environment of the main chain. We implement our design on a MacBook Pro with 4 cores, 2.2GHz Intel i7 and 16 GB RAM. Considering the instability of the blockchain network, we tested each function in the smart contract hundreds of times. The smart contract is written in Solidity and the part of participants' interaction with the smart contract is written in Truffle based on the Javascript and Web3.js.

\subsection{Implementation details}

Our design is based on commitment scheme, thus there are many waiting periods on the process. Meanwhile, the length of the waiting period is variable and is determined by the provider according to market conditions. Hence, we chose to test the time of each phase instead of the whole design. Specifically, there are four main phases in our design: bid commitment, bid open, auction phase and ladder payment. Especially, the Ethereum network is unable to afford a lot of computing. In the auction phase, we choose 5-20 users and a total of 5-10 resources.

\subsection{Experiments on Simulated Network}

In order to demonstrate the performance of each phase, we test each stage in the local simulation network \emph{Ganache}. \emph{Ganache} uses ethereumjs to simulate full client behaviors locally and mining is only related to the execution time of transactions without considering other complicated factors such as network latency. We test the four phases mentioned earlier as follow.

For the convenience of testing, we write the function of calculating the commitment in the smart contract, and send the result and the guaranty to the commitment function while getting the result. Then we use the promise structure in the Javascript to ensure that the commitment is on the chain before opening commitment. In the trade phase, we take $e$ equal to 5 and get five different confirmations. In the table $\ref{table1}$, we use $c_i$ to denote the $i$th confirmation while enumeration types are used to represent different states in smart contracts. Table $\ref{table1}$ presents the time overhead of the three phases, respectively. The open phase takes significantly more time than the commitment phase because of more computing work of the open phase. And the time cost for different confirmations is almost the same.

\begin{table}[ht]
  \caption{Time on \emph{Ganache}}
  \label{table1}
  \centering
  \begin{tabular}{|c|c|c|c|}
    \hline
    operation & time&operation &time\\
    \hline
    commitment &36ms&open &69ms\\
    \hline
    $c_1$&33ms&$c_2$&33ms\\
    \hline
    $c_3$&33ms&$c_4$&33ms\\
    \hline
    $c_5$&34ms&&\\
    \hline
  \end{tabular}
\end{table}

\begin{figure}
\centering
\includegraphics[scale=0.45]{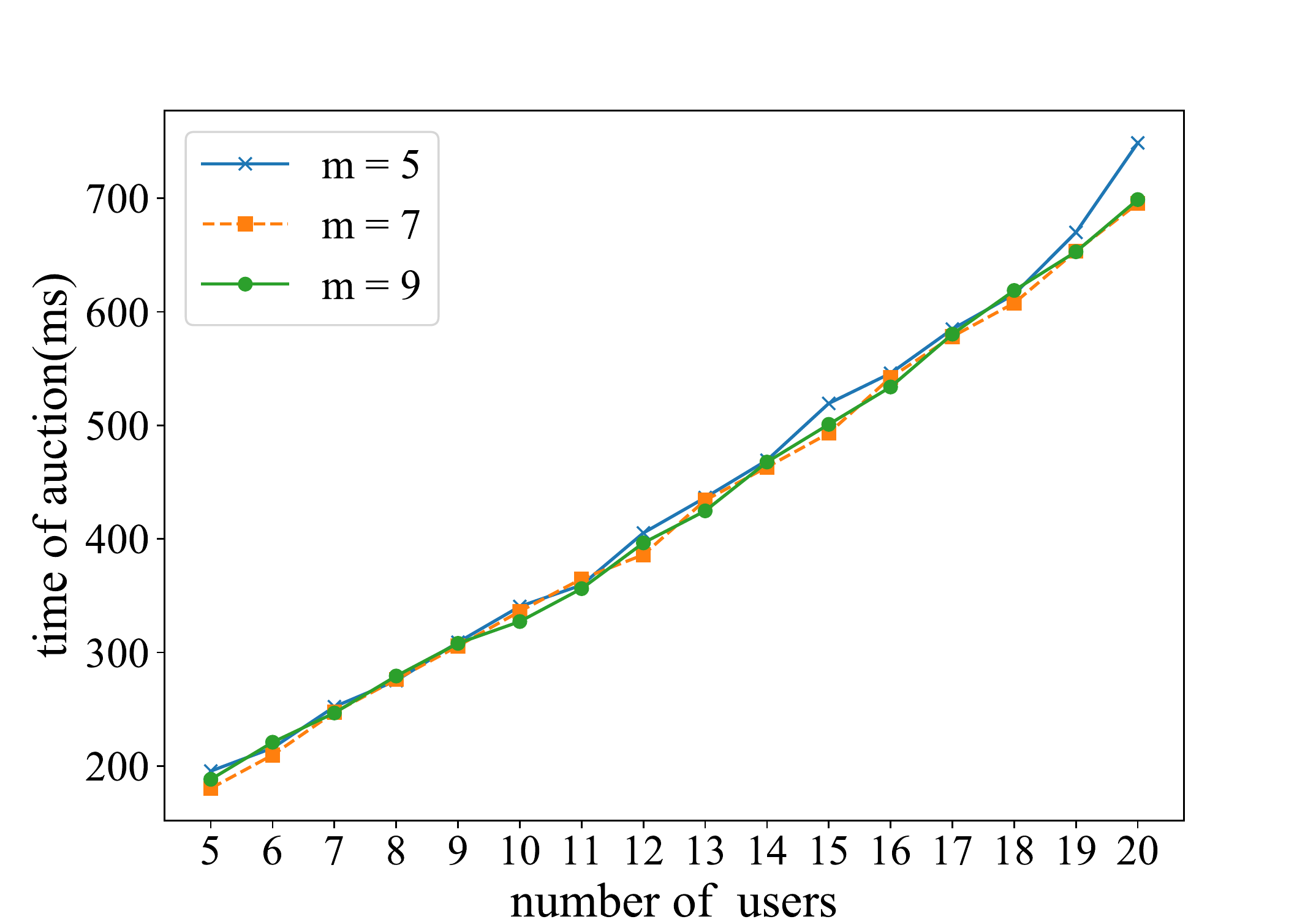}
\caption{The time of auction phase}
\label{fig:label7}
\end{figure}

Fig.~\ref{fig:label7} illustrates the auction time where the number of users ranges 5-20, and the numbers of VM types are 5,7,9. We test 40 times and average the times for each situation. We can observe that the number of VM types has a little impact on auction time overhead, while the number of users has a greater impact on the time overhead.

\subsection{Experiments on Test Network}
 In order to demonstrate the actual application of our design, we deploy each phase to the official test network \emph{Rinkeby} separately. As \cite{Ethereum2019} shows, the average mining time is 15$s$. In order to reduce the impact from network fluctuations or other negative effects such as broadcast latency, we test 100 times for each phase.

\begin{table}[ht]
  \caption{Time on \emph{Rinkeby}}
  \label{table2}
  \centering
  \begin{tabular}{|c|c|c|c|}
    \hline
    operation & time&operation &time\\
    \hline
    commitment &14990ms&open &15045ms\\
    \hline
    $c_1$&15001ms&$c_2$&14985ms\\
    \hline
    $c_3$&14998ms&$c_4$&15012ms\\
    \hline
    $c_5$&14827ms&&\\
    \hline
  \end{tabular}
\end{table}

\begin{figure}
\centering
\includegraphics[scale=0.45]{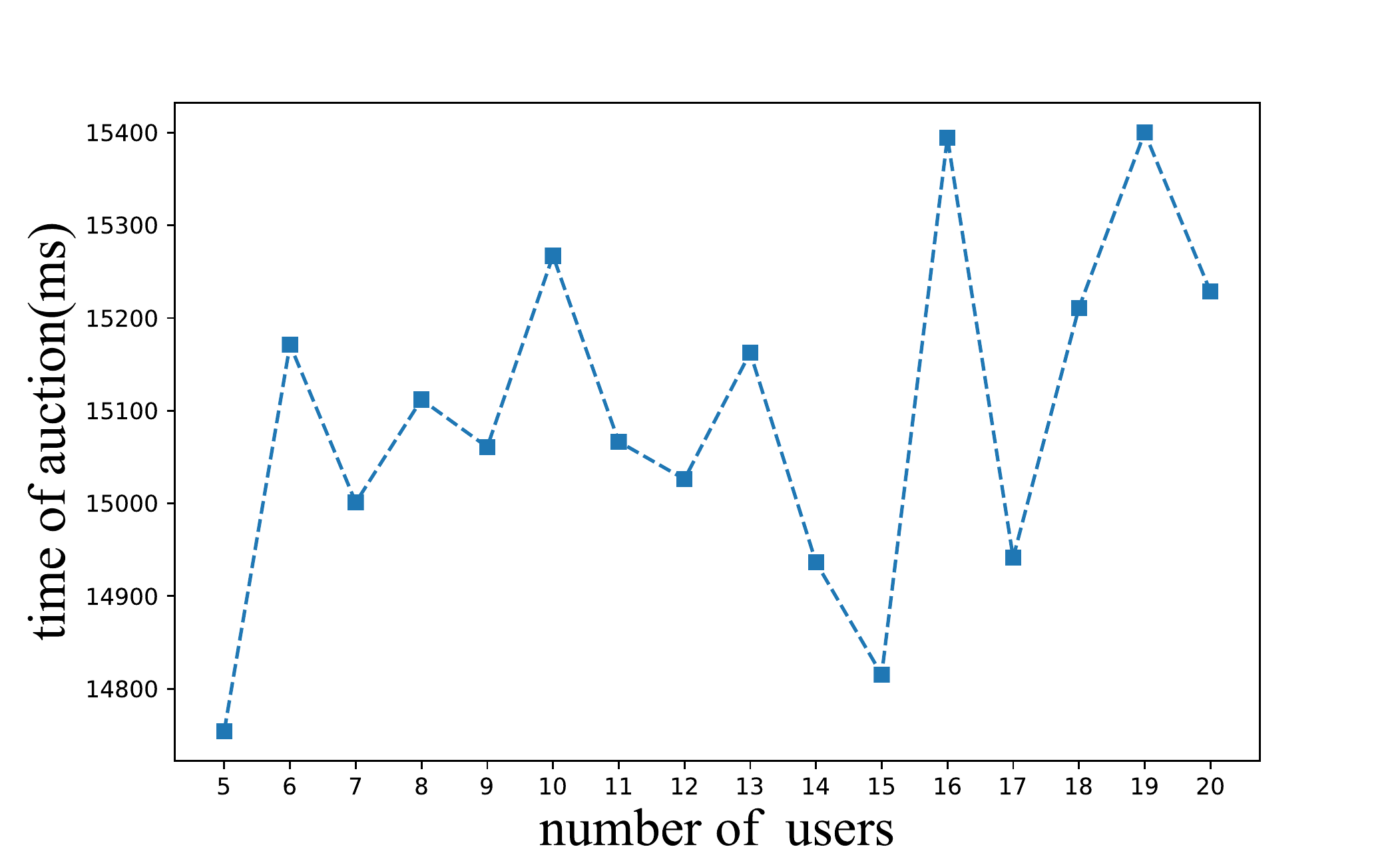}
\caption{The time of auction phase}
\label{fig:label8}
\end{figure}

The table~\ref{table2} indicates the average times on the \emph{Rinkeby} network. The average times of these three phases are around 15$s$. This means that all these phases can be chained in one block time. As Fig.~\ref{fig:label7} shows, the real auction time varies slightly with different number of VM types. Hence, we only compute when there are 5 VM types in the test network. Fig.~\ref{fig:label8} illustrates the time of auction phase. The auction times are distributed from 14754$ms$ to 15400$ms$, and thus they are around 15$s$. That is, the auction phase can be chained when the next block arrives, and the performance is acceptable.

\section{Conclusion}\label{sec:conclusion}
In this paper, we have proposed a fair auction and trade framework for cloud VM allocation based on blockchain. We have achieved both auction fairness and trade fairness by designing the \emph{state mechanism} based on commitment, digital currency and smart contract technologies. The auction fairness guarantees that users cannot alter their submitted bids and cannot quit the auction without any financial penalty, and that the malicious provider cannot alter the auction results. The trade fairness guarantees that participants cannot reject the trade without any financial penalty after the auction, and malicious users cannot get cloud VM instances without paying fees and the malicious provider cannot get fees without providing requested VM instances. And we further propose a fair trade without an adjudicator based on the relaxed trade fairness. We implement our design in Solidity, JavaScript and Web3.js, and experimentally demonstrate that the system performance is acceptable.

\bibliography{Reference}{}

\begin{thebibliography}{10}
\providecommand{\url}[1]{#1}
\csname url@samestyle\endcsname
\providecommand{\newblock}{\relax}
\providecommand{\bibinfo}[2]{#2}
\providecommand{\BIBentrySTDinterwordspacing}{\spaceskip=0pt\relax}
\providecommand{\BIBentryALTinterwordstretchfactor}{4}
\providecommand{\BIBentryALTinterwordspacing}{\spaceskip=\fontdimen2\font plus
\BIBentryALTinterwordstretchfactor\fontdimen3\font minus
  \fontdimen4\font\relax}
\providecommand{\BIBforeignlanguage}[2]{{%
\expandafter\ifx\csname l@#1\endcsname\relax
\typeout{** WARNING: IEEEtran.bst: No hyphenation pattern has been}%
\typeout{** loaded for the language `#1'. Using the pattern for}%
\typeout{** the default language instead.}%
\else
\language=\csname l@#1\endcsname
\fi
#2}}
\providecommand{\BIBdecl}{\relax}
\BIBdecl

\bibitem{samimi2016combinatorial}
P.~Samimi, Y.~Teimouri, and M.~Mukhtar, ``{A combinatorial double auction
  resource allocation model in cloud computing},'' \emph{Information Sciences},
  vol. 357, pp. 201--216, 2016.

\bibitem{baranwal2015fair}
G.~Baranwal and D.~P. Vidyarthi, ``{A fair multi-attribute combinatorial double
  auction model for resource allocation in cloud computing},'' \emph{Journal of
  systems and software}, vol. 108, pp. 60--76, 2015.

\bibitem{li2018online}
J.~Li, Y.~Zhu, J.~Yu, C.~Long, G.~Xue, and S.~Qian, ``{Online auction for IaaS
  clouds: Towards elastic user demands and weighted heterogeneous VMs},''
  \emph{IEEE Transactions on Parallel and Distributed Systems}, vol.~29, no.~9,
  pp. 2075--2089, 2018.

\bibitem{zaman2013combinatorial}
S.~Zaman and D.~Grosu, ``{Combinatorial auction-based allocation of virtual
  machine instances in clouds},'' \emph{Journal of Parallel and Distributed
  Computing}, vol.~73, no.~4, pp. 495--508, 2013.

\bibitem{bentov2014use}
I.~Bentov and R.~Kumaresan, ``{How to use bitcoin to design fair protocols},''
  in \emph{International Cryptology Conference}.\hskip 1em plus 0.5em minus
  0.4em\relax Springer, 2014, pp. 421--439.

\bibitem{andrychowicz2014secure}
M.~Andrychowicz, S.~Dziembowski, D.~Malinowski, and L.~Mazurek, ``{Secure
  multiparty computations on bitcoin},'' in \emph{2014 IEEE Symposium on
  Security and Privacy}.\hskip 1em plus 0.5em minus 0.4em\relax IEEE, 2014, pp.
  443--458.

\bibitem{kumaresan2014use}
R.~Kumaresan and I.~Bentov, ``{How to use bitcoin to incentivize correct
  computations},'' in \emph{Proceedings of the 2014 ACM SIGSAC Conference on
  Computer and Communications Security}.\hskip 1em plus 0.5em minus 0.4em\relax
  ACM, 2014, pp. 30--41.

\bibitem{zhang2018outsourcing}
Y.~Zhang, R.~Deng, X.~Liu, and D.~Zheng, ``{Outsourcing service fair payment
  based on blockchain and its applications in cloud computing},'' \emph{IEEE
  Transactions on Services Computing}, 2018.

\bibitem{zhang2018blockchain}
Y.~Zhang, R.~H. Deng, X.~Liu, and D.~Zheng, ``{Blockchain based efficient and
  robust fair payment for outsourcing services in cloud computing},''
  \emph{Information Sciences}, vol. 462, pp. 262--277, 2018.

\bibitem{hu2018searching}
S.~Hu, C.~Cai, Q.~Wang, C.~Wang, X.~Luo, and K.~Ren, ``{Searching an encrypted
  cloud meets blockchain: A decentralized, reliable and fair realization},'' in
  \emph{IEEE INFOCOM 2018-IEEE Conference on Computer Communications}.\hskip
  1em plus 0.5em minus 0.4em\relax IEEE, 2018, pp. 792--800.

\bibitem{delgado2017fair}
S.~Delgadosegura, C.~Perezsola, G.~Navarroarribas, and J.~Herrerajoancomarti,
  ``A fair protocol for data trading based on bitcoin transactions,''
  \emph{Future Generation Computer Systems}, vol. 2017, p. 1018, 2017.

\bibitem{huang2018bitcoin}
H.~Huang, X.~Chen, Q.~Wu, X.~Huang, and J.~Shen, ``{Bitcoin-based fair payments
  for outsourcing computations of fog devices},'' \emph{Future Generation
  Computer Systems}, vol.~78, pp. 850--858, 2018.

\bibitem{wang2012cloud}
Q.~Wang, K.~Ren, and X.~Meng, ``{When cloud meets ebay: Towards effective
  pricing for cloud computing},'' in \emph{2012 Proceedings IEEE
  INFOCOM}.\hskip 1em plus 0.5em minus 0.4em\relax IEEE, 2012, pp. 936--944.

\bibitem{nakamoto2008bitcoin}
S.~Nakamoto \emph{et~al.}, ``{Bitcoin: A peer-to-peer electronic cash
  system},'' 2008.

\bibitem{Gavin2014ethereum}
G.~Wood, \emph{{Ethereum: A secure decentralised generalised transaction
  ledger}}, 2014.

\bibitem{buterin2014next}
V.~Buterin and Others, ``{A next-generation smart contract and decentralized
  application platform},'' \emph{white paper}, 2014.

\bibitem{Amazon2018a}
\BIBentryALTinterwordspacing
Amazon, ``{Amazon EC2 spot instances},'' 2018. [Online]. Available:
  \url{http://aws.amazon.com/ec2/spot-instances/}
\BIBentrySTDinterwordspacing

\bibitem{wang2013reverse}
X.~Wang, J.~Sun, H.~Li, C.~Wu, and M.~Huang, ``{A reverse auction based
  allocation mechanism in the cloud computing environment},'' \emph{Applied
  Mathematics {\&} Information Sciences}, vol.~7, no.~1, pp. 75--84, 2013.

\bibitem{andrychowicz2014fair}
M.~Andrychowicz, S.~Dziembowski, D.~Malinowski, and L.~P. Mazurek, ``{Fair
  Two-Party Computations via Bitcoin Deposits},'' \emph{financial
  cryptography}, vol. 2013, pp. 105--121, 2014.

\bibitem{bentov2017instantaneous}
I.~Bentov, R.~Kumaresan, and A.~Miller, ``Instantaneous decentralized poker,''
  \emph{international conference on the theory and application of cryptology
  and information security}, vol. 2017, pp. 410--440, 2017.

\bibitem{Ethereum2019}
\BIBentryALTinterwordspacing
Ethereum, ``{Average mining time},'' 2019. [Online]. Available:
  \url{https://www.rinkeby.io/#stats}
\BIBentrySTDinterwordspacing

\end{thebibliography}
\bibliographystyle{IEEEtran}

\end{document}